\documentclass{jfm}
\usepackage{graphicx}
\usepackage{epstopdf, epsfig}
\usepackage{amssymb,amsmath}
\usepackage{epsfig}
\usepackage{color}
\usepackage{hyperref}

\newcommand\Ray{\mbox{\textit{Ra}}}  

\shorttitle{Inherent thermal convection in a granular gas inside a box}
\shortauthor{F. {Vega Reyes}, A. Puglisi, G. Pontuale and A. Gnoli}

\title{Inherent thermal convection in a granular gas inside a box under a gravity field}

\author{Francisco Vega Reyes\aff{1}
  \corresp{\email{fvega@unex.es}},
  Andrea Puglisi\aff{2}, Giorgio Pontuale\aff{2}
 \and Andrea Gnoli\aff{2}}

\affiliation{\aff{1}Departamento de F\'isica and Instituto de Computaci\'on Cient\'ifica Avanzada (ICCAEx), 06071 Badajoz, Spain
\aff{2}Istituto dei Sistemi Complessi, CNR and Dipartimento di Fisica, Universit\`a di Roma Sapienza, Piazzale Aldo Moro 2, 00185 Rome, Italy}

\begin{document}

\maketitle

\begin{abstract}
 We theoretically prove the existence in granular fluids of a thermal
 convection that is  inherent, in the sense that is always present and has
 no thermal gradient threshold (convection occurs for all finite values of the
 Rayleigh number). More specifically, we
 study a gas of inelastic smooth hard disks enclosed in a rectangular region under a
constant gravity field. The vertical  walls act as energy sinks (i.e., inelastic walls
 that are parallel to gravity) whereas the other two walls are perpendicular to gravity and
act as energy sources. We show that this convection is due to the combined action of dissipative lateral
walls and a volume force (in this case, gravitation). Hence, we call it
\textit{dissipative lateral walls convection}, DLWC. Our theory, that describes also
the limit case of elastic collisions, shows that inelastic particle collisions
enhance the DLWC. We perform our study via
numerical solutions (volume element method) of the corresponding hydrodynamic equations,
in an extended Boussinesq approximation. We show our theory describes the
essentials of the results for similar (but more complex) laboratory experiments. 
\end{abstract}


\section{Introduction}
\label{intro}

Fluid matter exhibits a remarkable tendency to build up patterns and organized
dynamical structures. Different stages can appear in a fluid system as we depart
further away from equilibrium \citep{G95,K01}: laminar convection ends up developing
turbulence \citep{B82} that produces different characteristic patterns
\citep{HD99} such as plumes, swirls, eddies, vortices \citep[the
  vortex inside Saturn's hexagon is a beautiful example of atmospheric
  stable vortex, see for instance the work by][]{G90}, and eventually,
spatio-temporal chaos \citep*{EMPE00}. Furthermore, extensive
spatio-temporal chaos may render back again equilibrium-like states at
larger time/length scales \citep{E00}.

Since the seminal works by \cite{B00} and \cite{R16}, the fluid flow
in closed circuits (\textit{convection}), caused by the presence of
temperature inhomogeneities, has likely been one of the most studied
problems in science \citep*[the works by][are good reviews on the
  subject]{CH93,BPA00,MWG06}. The phenomenon is well known to be ubiquitous
  in nature, including biological systems. But let us describe
it again, at its simplest: we consider a real,
experimental system where there is a gravity force, keeping a fluid
layer at rest at temperature $T$ and with two horizontal limiting
fluid surfaces. The upper one (in the sense of gravity) is either free
or in contact with a solid surface. Then, by means of some kind of
temperature source at higher temperature $T_0>T$, the fluid is heated
from below. The difference $T_0-T$ is gradually increased but the
fluid remains static. However, when a critical temperature gradient is
reached, fluid motion is set on, shaping regular patterns in all of
the fluid volume \citep{B00}.  According to classical
theory, the Rayleigh number  is the convection control parameter in this kind of
problem. This dimensionless parameter is usually defined as $\Ray\equiv \alpha
\Delta T gh^3/\kappa\nu$, where $\alpha\equiv (1/V)(\partial T/\partial V)_p$
 is the fluid expansion ($V$ is the fluid volume and $T$ and $p$ temperature and
 hydrostatic pressure respectively) coefficient, $\Delta T$ is the boundaries temperature difference, $g$
is gravity acceleration, $h$ is the system width and $\kappa$ and $\nu$ are the
thermal conductivity and kinematic viscosity transport coefficients, respectively. In fact, linear
theory predicts a critical value $\Ray_c = 657.5$ and $\Ray_c = 1708$ for the free
surface and the closed on top system cases respectively \citep{MWG06}. The theoretical treatment by \cite{R16}
relied on the work by \cite{B03} \citep[see also the book edited by][for a
more recent review]{MWG06}, who defined the relevant
contributions for this convection in the fluid balance equations. These balance
equations, as it is known, were worked out by C.-L. Navier
and G. G. Stokes only a few years before \citep{B67}.

Let us recall also that a more generic concept of fluid involves also
systems where the particles are not necessarily \textit{microscopic};
i.e., the particles can be \textit{macroscopic}, when their typical size is greater than $1~\mu\mathrm{m}$ \citep{B54}. In fact, the dynamical
properties of a set of rigid macroscopic particles in a high state of agitation
was elucidated as a subject of the theory of fluids a long time ago by
\cite{R85}. However, for
macroscopic particles the kinetic energy will be partially
transferred, upon collision, to the lower (smaller length scales)
dynamics levels, never coming back to the upper granular level. For
instance, it may be transferred into thermal movement of the molecules
that are the constituents of the disk material \citep*{AFP13}. Thus,
unless the system gets an energy input from some kind of source, it
will evolve by continuously decreasing its total kinetic energy;
i.e. lowering the system \textit{granular temperature}
\citep{K79}. This temperature decay rate was calculated, for a
homogeneous and low density granular system (i.e., a homogeneous
\textit{granular gas}), by \cite{H83}.


Nevertheless, when excited by some persistent external action, stable
granular gas systems are found spontaneously in nature, for instance
in sand storms \citep{B54}, and also in laboratory experiments, where
air flow \citep*{LBLG00} or mechanical vibration \citep*{OU98,PG12}
may be used as energy inputs. Under these conditions, the granular gas
can develop steady laminar flows \citep{VU09}. Unfortunately, the hydrodynamics
of granular fluids is not in the same stage of development as it is for
molecular fluids \citep{P15}. For instance, the corresponding hydrodynamic theory for thermal convection
in a granular gas was developed only very recently
\citep[see for instance the work by][]{KM03} and only in the case of the
academic problem of horizontal (or inclined)
infinite walls. But, obviously real systems are finite. In the case of a gas
heated from two horizontal walls the simplest finite configuration considered is
when the system is closed by adding vertical lateral walls. As it is known, finite size effects have an impact in both the critical Rayleigh number and
the convection scenario in a molecular fluid changes if a cold vertical
wall is present \citep*{D77,HW77,MWG06}.

The lack of a theoretical analysis on the effects of finite size
systems in  granular convection theory \citep[see for instance the works by][and
others]{FP03,KM03,B10,EWea07,EMea10} may have hindered and/or rendered
not possible a complete interpretation of a part of the previous results on granular
dynamics laboratory
experiments. Furthermore, as we
will see in some experimental works, the observed granular convection
\cite*{WHP01,WHP01E,RSGC05,EMea10,WRP13,PGVP16} should be either exclusively or partly due to
sidewall energy sink, and not of Rayleigh-B\'enard type. And of course, although the no-sidewalls theoretical
approach may be accurate when bulk convection is present \citep{KM03}, in the cases where the
convection is caused only due to sidewalls energy sink we may expect
the convection properties to be very different.
Notice that in an enclosed granular gas the
lateral energy sink should always be present, since wall-particle
collisions are inherently inelastic. This implies that the present analysis
should be relevant for many granular convection experiments. Furthermore, as we
will see, lateral wall effects are also
more substantial for the granular gas than for the molecular fluid. 


\section{Description of the system and the problem}

 Let us consider a system consisting of a large set of circular
 particles (disks) in a two-dimensional (2D) system. The particles are
 identical inelastic smooth hard disks with mass $m$ and diameter $\sigma$. The
 hard collision model works reasonably well at an experimental level for a
 variety of materials \citep*{FLCA94}. We
 use here this model in the smooth particle approximation; i.e., we neglect the
 effects of sliding and friction in the collision.  Under the smooth hard particle model for
 collisions, the fraction of kinetic energy loss after collision is
 characterized by a constant parameter called the coefficient of normal
 restitution $\alpha$, not to be confused with the expansion coefficient, usually
 denoted also as $\alpha$.
 
 \begin{equation}
   \boldsymbol{n\cdot v'}_{12} = -\alpha \boldsymbol{n\cdot v}_{12},
   \label{collisional_rule} 
 \end{equation} where $\boldsymbol{v}_{12}, \boldsymbol{v'}_{12}$ are the
 collision pair contact
 velocities before and after collision, respectively \citep{FLCA94}.

 The system is under the action of a constant gravitational
 field $\boldsymbol{g}=-g\boldsymbol{\hat e}_y$. We will also assume that the system has low particle
 density ($n$) everywhere at all times.  Therefore, our
   fluid is a granular gas. Collisions are
 instantaneous \citep[in the sense that the contact time
   is very short compared to the average time between
   collisions,][]{CC70} and occur only  between two particles. Since particles
   are inelastic, our theory should take into account the inelastic cooling
   term.

\begin{figure}
 \centerline{\includegraphics[width=.75 \columnwidth]{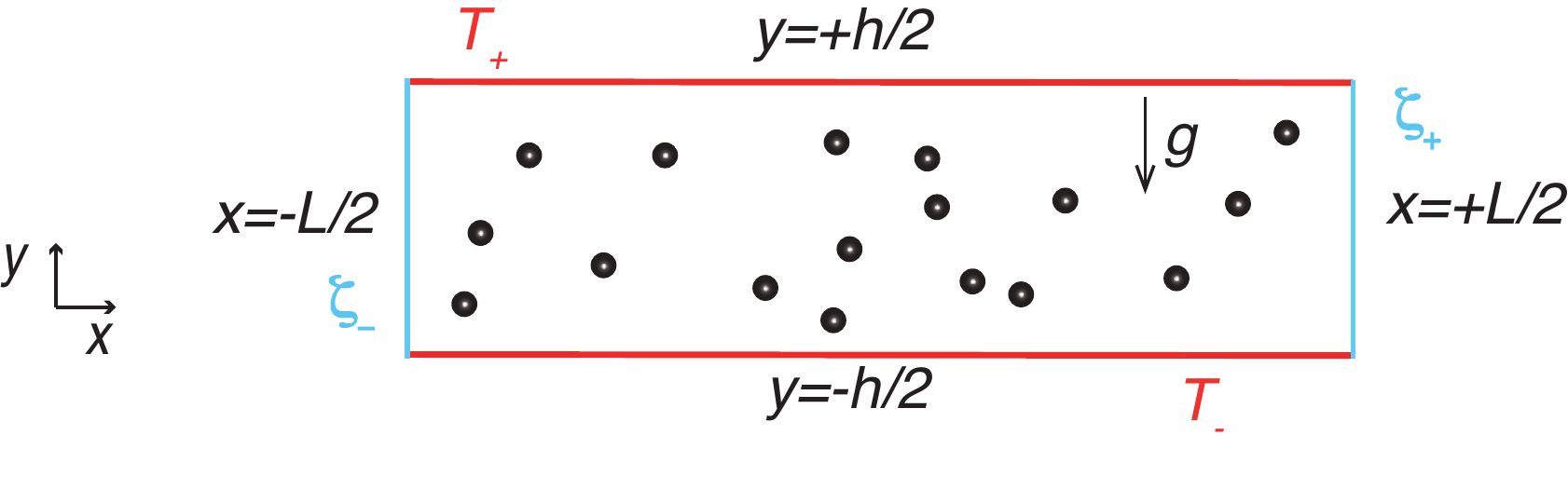}}
 \caption{Simple sketch of the system. The system is heated
   from the horizontal walls (at $y=\pm h/2$). If the lateral walls (at $x=\pm L/2$)
   act as energy surface sinks, bi-dimensional flow occurs.}
 \label{sketch}
 \end{figure}

   The system is provided  with either two (top and bottom) or just one (bottom, in
   the sense of gravity)
   \textit{horizontal walls} (i.e.,
   perpendicular to gravity), these being provided with energy
 sources. In addition, our granular gas is caged in a finite
    rectangular region by two inert \textit{vertical walls} (we call them
    lateral walls, or sidewalls).  Sidewalls-particle
    collisions are inherently inelastic, the degree of inelasticity of these
    collisions being characterized by a coefficient of normal restitution
    $\alpha_w$ that is in general different from the one for particle-particle collisions, $\alpha$. See
    Figure~\ref{sketch} for a graphical description.

We denote the single particle
velocity distribution function as $f(\boldsymbol{r},\boldsymbol{v}|t)$
with $\boldsymbol{r}, \boldsymbol{v}$ being the particle position and
velocity, respectively functions of time ($t$). The first three
velocity moments of the distribution function
$n(\boldsymbol{r},t)=\int\mathrm{d}\boldsymbol{v}f(\boldsymbol{r},\boldsymbol{v}|t)$,
$\boldsymbol{u}(\boldsymbol{r},t)=(1/n)\int{\mathrm{d}\boldsymbol{v}f(\boldsymbol{r},\boldsymbol{v}|t)}\boldsymbol{v}$
,
$T(\boldsymbol{r},t)=(1/dn)\int{\mathrm{d}\boldsymbol{v}f(\boldsymbol{r},\boldsymbol{v}|t)}mV^2$,
define the average fields particle density ($n$), flow velocity
($\boldsymbol{u}$) and  temperature ($T$), respectively. Here,  $\boldsymbol{V}=\boldsymbol{v}-\boldsymbol{u}$ and $d$ is the
system dimension. In this work we consider only $d=2$ (and that is why the
particles are necessarily flat).

For a granular gas, molecular chaos
(i.e., particle velocities are not statistically correlated) also occurs in
most practical situations \citep*{PEU02,BO07}. Therefore, the kinetic
Boltzmann equation \citep{CC70}, may also be used to describe granular
gases \citep*{D01,BDKS98}. The general balance equations that follow from the inelastic Boltzmann equation
 have the same form that for molecular fluid except for the additional inelastic
 cooling term arising in the energy equation. They have the following form \citep{BDKS98,SG98}

 \begin{equation}
 {D_tn}=-n\nabla\cdot\boldsymbol{u} , \quad {D_t\boldsymbol{u}}=-\frac{1}{mn}\boldsymbol\nabla\cdot{\mathsfbi{P}}+\boldsymbol{g},
\quad  {D_tT}+\zeta T=-\frac{2}{dn}\left(\mathsfbi{P}\boldsymbol{:\nabla}\boldsymbol{u}+\boldsymbol\nabla\cdot{\boldsymbol q}\right).
 \label{bal_eq}
 \end{equation} In the above equations, $D_t\equiv \partial_t+\boldsymbol{u}\cdot\boldsymbol\nabla$ is the material derivative \citep{B67}, and $\mathsfbi{P}, \boldsymbol{q}$ are the moment and energy fluxes (stress tensor and heat flux), defined respectively by $\mathsfbi{P}=m\int \mathrm{d}\boldsymbol{v}\,\boldsymbol{V}\boldsymbol{V}f(\boldsymbol{v})$ and
$\boldsymbol{q}=(m/2)\int \mathrm{d}\boldsymbol{v}\,V^2\boldsymbol{V}f(\boldsymbol{v})$.
 
As we said, notice the new term $\zeta T$ in the energy equation of
\eqref{bal_eq}, where $\zeta$ represents the rate of kinetic energy loss, and is
usually called \textit{inelastic cooling rate}. Accurate expressions of the
cooling rate and the inelastic Boltzmann equation are well known and may be
found elsewhere \citep[we use here the one worked out by][]{BDKS98}.


 Let us note also that the set of balance equations \eqref{bal_eq} is exact and
 always valid. However, in order to close
 the system of equations, we need to express the fluxes
 $\mathsfbi{P}, \boldsymbol{q}$ and the cooling rate $\zeta$ as functions of the
 average fields $n, \boldsymbol{u}, T$. Hydrostatic pressure field $p$ is
 defined by the equation of state for an ideal gas: $p=nT$. Starting out of the
 kinetic equation for the gas, this can only be done if the distribution
 function spatio-temporal dependence can be expressed through a functional
 dependence on the average fields; i.e., if the gas is in a \textit{normal
   state}  \citep{H12}, this only being true if the spatial gradients vary over
 distances greater  than the mean free path (that is, the characteristic
 microscopic scale). In this case, it is usually said that there is \textit{scale separation} \citep{G03}. Henceforth, we assume this scale separation occurs at all situations considered for this work \citep[see the reference by][for more detail about the conditions for accuracy of this assumption in steady granular gas flows]{VU09}.
 
The boundary conditions come from usual forms for temperature sources,
no-slip velocity \citep{VSG10}, and controlled pressure at the boundaries:
$T(y=+h/2)=T_+$, (substituted by $[\partial T/\partial y =0]_{y=+h/2}$, with
$h\gg L$, in the case of an open on top system), 
$T(y=-h/2)=T_-$, $\boldsymbol{u}(x=\pm L/2)=\boldsymbol{u}(y=\pm
h/2)=\boldsymbol{0}$, $p(y=-h/2)=p_0$. For an enclosed granular gas we also
necessarily need to consider the dissipation at the lateral walls, as we previously
explained. A condition for the horizontal derivative of temperature would suffice to
account for an energy sink at the side walls \citep{HW77}. However, taking into
account that the energy sink comes from wall-particle collision inelasticity, then it
is more appropriate a horizontal heat flux that is proportional to
lateral wall-disks collisions degree of inelasticity
inelasticity, $\propto (1-\alpha_w^2)$ \citep{JJ87},
\begin{equation}
q_x(x=\pm L/2) = \mathcal{A}(\alpha_w) \left[pT^{3/2}\right]_{x=\pm L/2},
\label{DLW_bc}
\end{equation} where $\mathcal{A}(\alpha_w)=(\pi/2)m(1-\alpha_w^2)$ is given by
the dilute limit of the corresponding expression in the work by
\cite*{NAAJS99}. (Additionally, condition $p(x=-L/2)=p(x=+L/2)$ is also
implicitly used, since in this work we only consider identical lateral walls at
both sides). The detailed balance of fluxes across the boundaries is beyond the
scope of this work, but for a more detailed analysis on realistic boundary conditions the
reader may refer to the work by \cite*{NAAJS99}.

 \subsection{Navier-Stokes equations and transport coefficients}

 We will assume that the spatial gradients are sufficiently small, which is true for steady laminar flows near the elastic limit \citep[][]{VU09}. Therefore, we use the Navier-Stokes  constitutive relations for the fluxes \cite{BDKS98,BC01}

 \begin{align}
\mathsfbi{P}&=p\mathsfbi{I}-\eta\left[\boldsymbol\nabla\boldsymbol{u}+\boldsymbol\nabla\boldsymbol{u}^\dagger-\frac{2}{d}\mathsfbi{I}(\boldsymbol\nabla\cdot\boldsymbol{u})\right],	\\
 \boldsymbol{q}&=-\kappa\boldsymbol\nabla T-\mu\boldsymbol\nabla n.
 \label{fluxes_NS}
 \end{align}

 In \eqref{fluxes_NS} we can find the transport coefficients: $\eta$
 (viscosity),  $\kappa$ (thermal conductivity), and $\mu$ (thermal
 diffusivity). Notice that $\mu$ is a new coefficient that arises from inelastic
 particle collisions \cite{BDKS98}. The transport coefficients for the granular
 gas have been calculated by several authors, with some variations
in the theoretical approach. For instance, \cite{SG98} performed a
Chapman-Enskog-like power series expansion in terms of both the spatial
gradients and inelasticity, up to Burnett order, but limited to quasi-elastic particles whereas \cite{BDKS98} perform the expansion only in the spatial gradients (and the theory is formally valid for all values of inelasticity). Previous works, like the work by \cite{JS83} and by \cite*{LSJC84} obtain the granular gas transport coefficients only in the quasi-elastic limit. These theories will actually yield indistinguishable values of the Navier-Stokes transport coefficients for nearly elastic particles (the case of our interest in the present work).





The essential point to our problem is the scaling with temperature $T$ and particle density $n$. This scaling for hard particles is \citep{CC70,BDKS98}

 \begin{equation}
 \eta=\eta_0^*T^{1/2}\: , \quad \kappa=\kappa_0^*T^{1/2} \: ,
 \quad \mu=\mu_0^*\frac{T^{3/2}}{n} \: , \quad \zeta=\zeta_0^*\frac{p}{T^{1/2}} \:.
 \label{coefs0}
 \end{equation} 

The values of the coefficients for disks are $\eta^*_0\equiv\eta^*(\alpha)\sqrt{m}/(2\sigma\sqrt{\pi}),
\kappa^*_0\equiv 2\kappa^*(\alpha)/(\sqrt{\pi m\sigma}), \mu^*_0\equiv 2\mu^*(\alpha)/(\sqrt{\pi m\sigma}),
\zeta^*_0\equiv(2\sigma\sqrt{\pi/m})\zeta^*(\alpha)$ the expressions dimensionless functions
that depend on the coefficient of restitution can be found in the work by
\cite{BC01}. We write them here for completion

\begin{subequations}
\begin{align}
  & \eta^*(\alpha)=\left[\nu^*_1(\alpha)-\frac{\zeta^*(\alpha)}{2}\right]^{-1}, \\
  & \kappa^*(\alpha)=\left[\nu^*_2(\alpha)- \frac{2d}{d-1}\zeta^*(\alpha)\right],
  \\
  &
    \mu^*(\alpha)=2\zeta^*(\alpha)\left[\kappa^*(\alpha+\frac{(d-1)c^*(\alpha)}{2d\zeta^*(\alpha)})\right],
  \\
  & \zeta(\alpha^*)= \frac{2+d}{4d}(1-\alpha^2)\left[1+\frac{3}{32}c^*(\alpha)\right],
\end{align}
  \label{coefficients}
\end{subequations} where

\begin{subequations}
\begin{align}
  &
    \nu^*_1(\alpha)=\frac{(3-3\alpha+2d)(1+\alpha)}{4d}\left[1-\frac{1}{64}c^*(\alpha)\right],\\
  & \nu^*_2(\alpha)= \frac{1+\alpha}{d-1}\left[ \frac{d-1}{2} +
    \frac{3(d+8)(1-\alpha)}{16}+\frac{4+5d-3(4-d)\alpha}{1024}c^*(\alpha)\right],\\
  & c^*(\alpha)= \frac{32(1-\alpha)(1-2\alpha^2)}{9+24d+(8d-41)\alpha+30\alpha^2(1-\alpha)},
\end{align}
\label{coefficients_functions}
\end{subequations} and in our system $d=2$ since we deal with disks.






\subsection{The heated granular gas: convective base state}
\label{convective}

First, we revisit the general argument which states that a hydrostatic
state is impossible when a temperature gradient is assumed in the
horizontal (transverse to gravity) direction, as it is in the case of
dissipative lateral walls \cite{PGVP16}; i.e., when $T = T(x,y)$.  Momentum balance in Eq.~\eqref{bal_eq}
supplemented by ideal equation of state, in the absence of macroscopic
flow (hydrostatic) states that
\begin{subequations}
\label{2d_balance}
\begin{align}
\partial_x p=\partial_x(n(x,y)T(x,y))=0\\
\partial_y p=\partial_y (n(x,y)T(x,y))=-mgn(x,y).
\end{align}
\end{subequations}
The first equation yields $p(x,y) \equiv p(y)$, which, used in the second equation,
sets $n(x,y) \equiv n(y)$ and, using this back in the first equation above, we get $T(x,y)\equiv T(y)$; i.e.,
$\partial T/\partial x = 0$. This is in contradiction with the horizontal
temperature gradient assumed above. Therefore, the simple hydrostatic system of
equations is not compatible with the condition $\partial T/\partial x\neq 0$,
originated by the energy sinks at the side walls.

We must conclude that a hydrostatic state in the presence of DLW and
gravity is not possible. Thus, since $\boldsymbol{u}\neq 0$ a flow must always be present, even at
infinitesimal values of the Rayleigh number; i.e. there is no hydrostatic
solution even if $\mathrm{Ra}\to 0$.
In the terminology of previous
bibliography on molecular fluids enclosed by dissipative sidewalls,
\textit{a smooth
  transition occurs so that the concept of a critical Rayleigh number is no
  longer tenable} \citep{HW77}.

\section{Extended Boussinesq-like approximation for a granular gas}
\label{boussinesq}


However, our previous analysis does not
explain why a convection in the form of that observed in the experiments
appears. Indeed, our analysis only implies that there is never a hydrostatic
solution: it does not lead necessarily to a flow with one convection cell next
to each inelastic wall, nor to a convection-free region for points sufficiently
far away from the side walls, as seen in the experiments
\citep{WHP01,WHP01E,RSGC05,EMea10,WRP13,PGVP16}. Furthermore,
we also need to discard if other properties, other than sidewalls inelasticity,  present in the experimental
system (such as particle-bottom plate friction) are important for the appearance
of DLWC in the granular gas. 

Therefore, we need to analyze in \S \ref{boussinesq} the minimal system of
differential equations that derives from the general balance equations and that
is able to reproduce a convection with the characteristics of the one observed
in the experiments. Once it
is numerically solved, we will be able to describe in detail the main features
of the non-hydrostatic base state in our system.

According to both experiments and computer simulations (molecular dynamics)
results, this convection would only show one cell per inert wall, independently of
thermal gradients strength and system size \cite{PGVP16}. The flow in the bulk
of the fluid, for wide systems, appears to be zero or negligible. This is
analogous to the result for molecular fluids where the sidewalls effects are
important \citep{HW77}. Moreover, the presence of sidewalls introduces two
important but different in origin
effects. The first one arises from finite size effects alone and it shows up in
a lower critical Rayleigh number for B\'enard's convection, even if the lateral
walls are perfectly insulating (i.e., even if they do not convey a lateral energy flux). It is
impossible to escape this effect both in molecular \citep[as described in the
work by][]{HW77} and granular fluids when enclosed by lateral walls. The second
effect comes properly from energy dissipation at the lateral walls and as we saw
produces a non-hydrostatic steady state by default.

We are specifically interested in this second effect that arises only with
dissipative lateral walls and leave for future work the study of the first
effect (that should lead eventually to consider more appropriate theoretical
criteria when comparing with experimental results for the classical bulk
granular convection). Therefore, it is our
aim to elucidate the minimal theoretical framework that is able to take into account the important
experimental evidence of the DLWC in granular gases.

For our theoretical description, let us use as reference units: particle mass $m$ for mass, particle diameter $\sigma$ for length, thermal velocity at the base $v_0=(4\pi 
T_0/m)^{1/2}$ for velocity, pressure at the base $p_0=n_0T_0$ and $\sigma/v_0$ for time, where $T_0, n_0$ are arbitrary values of temperature and particle density, respectively. 

A common situation for thermal convection is that all density derivatives are
negligible, except for the spatial dependence of density that is coming from gravity, which appears in the momentum balance equation. This happens when the variation of mechanical energy is small compared to the variation of thermal energy \citep{GG76} and leads to the Boussinesq equations \citep{B78,C81}. This is always true in our system if the reduced gravitational acceleration fulfills $g\sigma/v_0^2\ll 1$. Thus, we restrict our analysis to small values of $g$. Taking this into account in the mass balance equation in \eqref{bal_eq}, immediately yields $\partial u_x/\partial x+ \partial u_y/\partial y=0$. 
Moreover, for weak  convection (as it is the case of our experimental results) we can
neglect the advection (nonlinear) terms that emerge in the balance equations
\eqref{bal_eq} \citep{B78}.

Incorporating these approaches into the other balance
equations in \eqref{bal_eq} and for our system geometry (see figure \ref{sketch}),
and with our reference units, we get the following dimensionless Boussinesq equations
for weak convection in the granular gas


\begin{align}
  & \eta^*(\alpha)\frac{\partial}{\partial y}\left[\sqrt{ T}\left(\frac{\partial u_x}{\partial y}+\frac{\partial u_y}{\partial x}\right)\right]+
2 \eta^*(\alpha)\frac{\partial}{\partial x}\left[\sqrt{ T}\frac{\partial u_x}{\partial x}\right]-\frac{\partial (n T)}{\partial x} = 0,  \label{linear_boussinesq_ux}\\
  & \eta^*(\alpha)\frac{\partial}{\partial x}\left[\sqrt{ T}\left(\frac{\partial u_x}{\partial y} + \frac{\partial u_y}{\partial x}\right)\right]+2 \eta^*(\alpha)\frac{\partial}{\partial y}\left[\sqrt{ T}\frac{\partial u_y}{\partial y}\right]-\frac{\partial ( n T)}{\partial y} - n g^* = 0,  \label{linear_boussinesq_uy} \\
  &  n^2 T^{3/2}\zeta^*(\alpha)=  \frac{\kappa^*(\alpha)}{\pi}\left[ \frac{\partial}{\partial x}\left(\sqrt{ T} \frac{\partial T}{\partial x}\right)+\frac{\partial}{\partial y}\left(\sqrt{ T} \frac{\partial T}{\partial y}\right)\right], \label{linear_boussinesq_T}
\end{align} with $g^*=4\pi g$. In fact, our Boussinesq-extended approximation 
includes additional terms with respect to the classical Boussinesq approximation used
for thermal convection, since we do not neglect the temperature dependence of the transport coefficients, which results in
$\sqrt{T}$ factors inside the bracket terms in equations
\eqref{linear_boussinesq_ux}-\eqref{linear_boussinesq_T}. In a previous work we
noticed that these temperature factors are relevant for important properties of the steady
profiles of the granular temperature, such as the curvature \citep{VU09}. Notice
we also keep the density derivatives in
\eqref{linear_boussinesq_ux}-\eqref{linear_boussinesq_uy}, since the granular
gas is highly compressible. For this reason, we do not strictly consider density
to be constant in the mass balance equation; we neglect density variations along
the flow field lines instead, while keeping density variations in the momentum
balance equation. The corresponding dimensionless forms of the boundary
conditions would be: $T(y=+h/2)=T_+/(mv_0^2/2)$, ($[\partial T/\partial y ]_{y=+h/2}=0$ for
an open on top system), 
$T(y=-h/2)=1/\sqrt{4\pi}$, $\boldsymbol{u}(x=\pm L/2)=\boldsymbol{u}(y=\pm
h/2)=\boldsymbol{0}$, $p(y=-h/2)=1$, plus for the lateral walls dissipation
condition $q_x(x=\pm L/2) = \mathcal{A}(\alpha_w) \left[pT^{3/2}\right]_{x=\pm L/2}$. 


\subsection{Numerical solution and comparison with 
experiments}

 \begin{figure}
 \vspace{8pt}
 \centerline{
 \includegraphics[height=4.0cm]{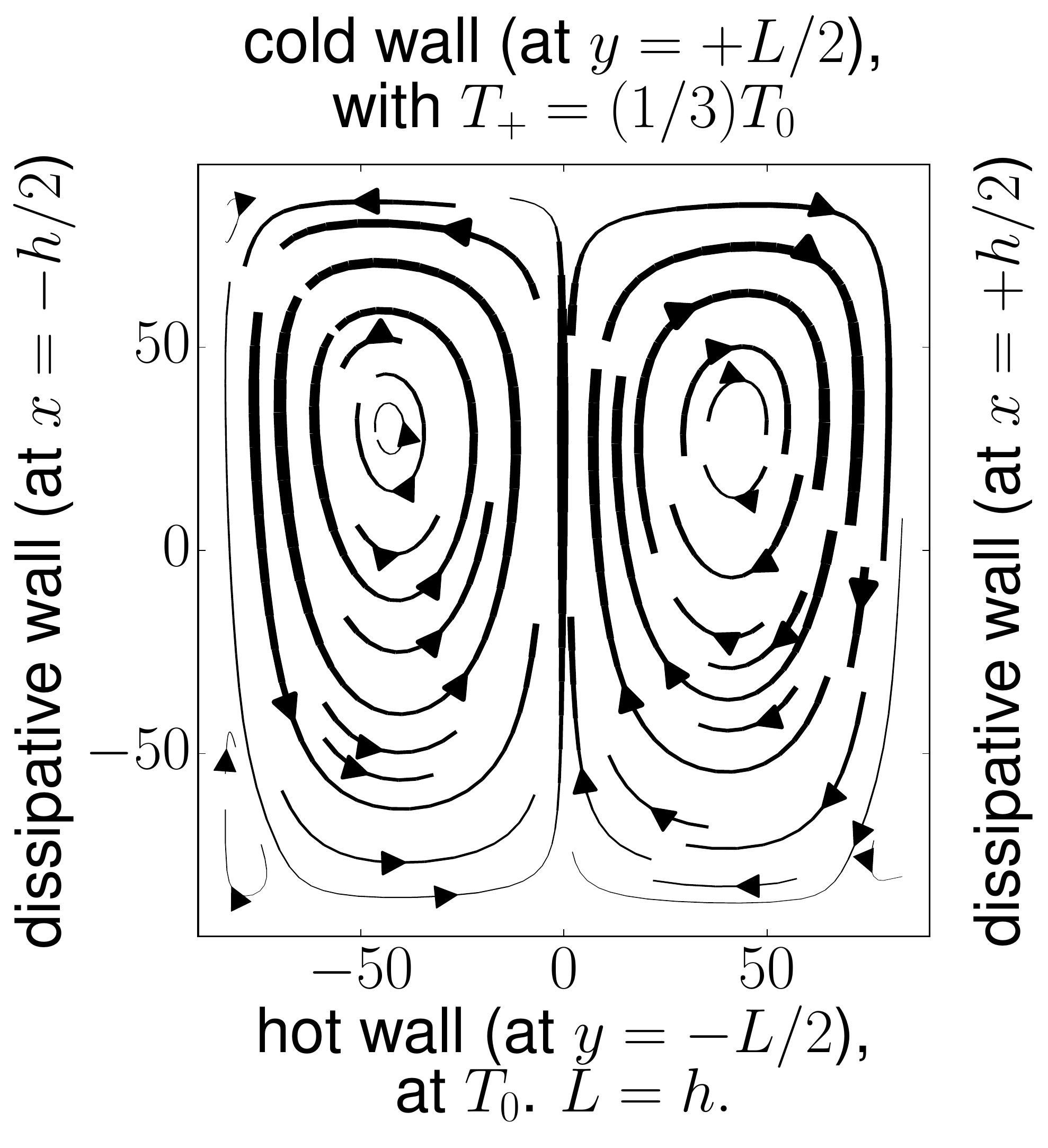} \quad
 \includegraphics[height=3.5cm]{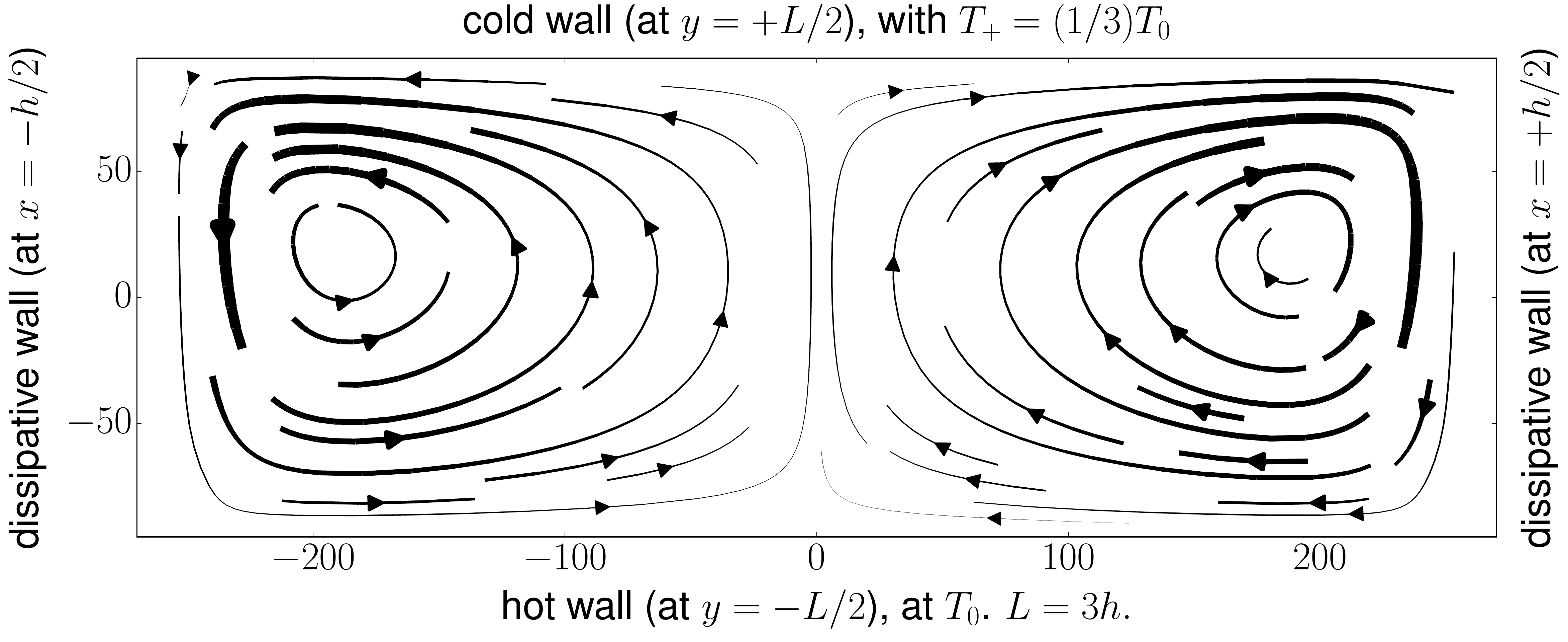}  \vspace{8pt}\\
 }
 \centerline{\includegraphics[height=3.5cm]{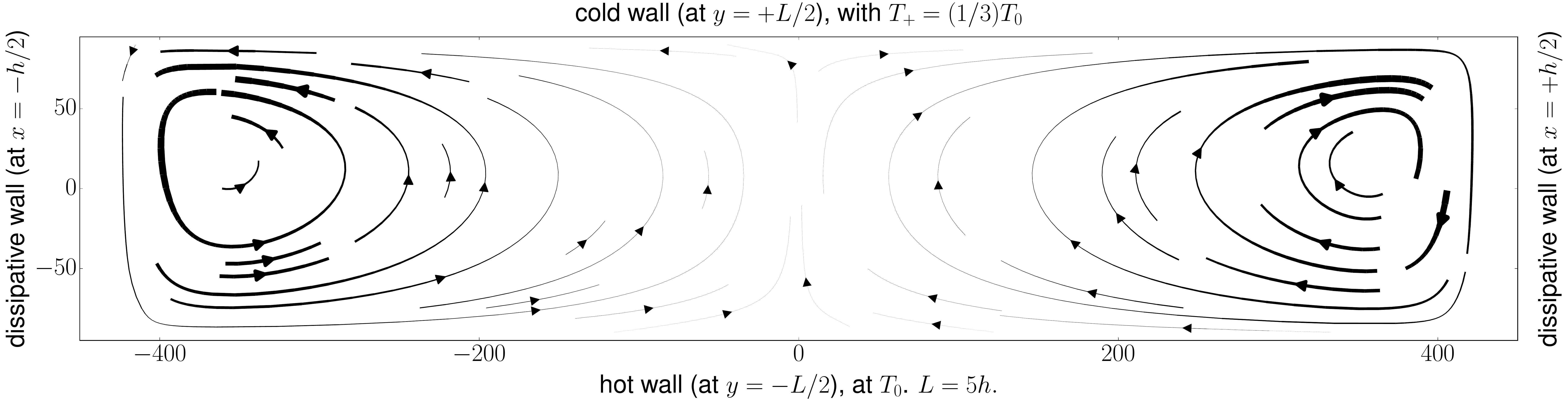} }
  \caption{DLW convection for systems with a top wall and different widths. Left
    to right  top to bottom: $L=h$, $L=3h$, $L=5h$. In our dimensionless units:
    system height $h=170$. In each plot, thickest stream lines correspond to
    $u_0 = (u_x^2+u_y^2)^{1/2} = 3.11\times 10^{-3}$ (in our dimensionless
    units). The other relevant parameters in this figure are: $T_0=3T_+$,
    $g=0.002\,g_0$, $\alpha=0.9$ (for both particle-particle and
    wall-particle collisions).}
 \label{fig_wide}
 \end{figure}

 In order to numerically solve the equations
 \eqref{linear_boussinesq_ux}-\eqref{linear_boussinesq_T}, we used the finite volume
 method. For this, we wrote a code using the SIMPLE algorithm \citep[in order to avoid numerical decoupling of the pressure field][]{FP02} and the FiPy differential equation package with the PySparse solver \citep*{GWW09}. We have seen (see figure 3) that the agreement with experiment and MD simulations is qualitatively very good. Also, all major properties of the flow are reproduced in the numerical solution obtained from the Boussinesq approximation. 

 Figure \ref{fig_wide}, that corresponds to systems provided with a top wall,
 clearly demonstrates that wider systems do not display more convection
 cells. The
 number of cells remains always one per dissipative wall. We also notice that
 the flow is upwards in the outer part of the cells (towards the system center)
 and downwards next to the lateral walls. It is interesting to note also that,
 according to numerical results, the convection speed $u_0=(u_x^2+u_y^2)^{1/2}$
 reaches roughly the same maximum value when varying system
 thickness. Furthermore, if we define the cell size as the horizontal distance
 between the upwards and downwards streams (see Figure 2) points with maximum convection speed, then we see that the size of the cells remain constant. The exception is for systems thinner than twice the cell size, in which case the cells squeeze each other (panel 1 in Figure \ref{fig_wide}). All of these results show the peculiarities of the DLW convection with respect to Rayleigh-B\'enard convection and coincide with the experimental behaviour previously detected \citep{PGVP16}.

\begin{figure}
  \begin{center}
  \includegraphics[width=3.5cm,height=4.1cm]{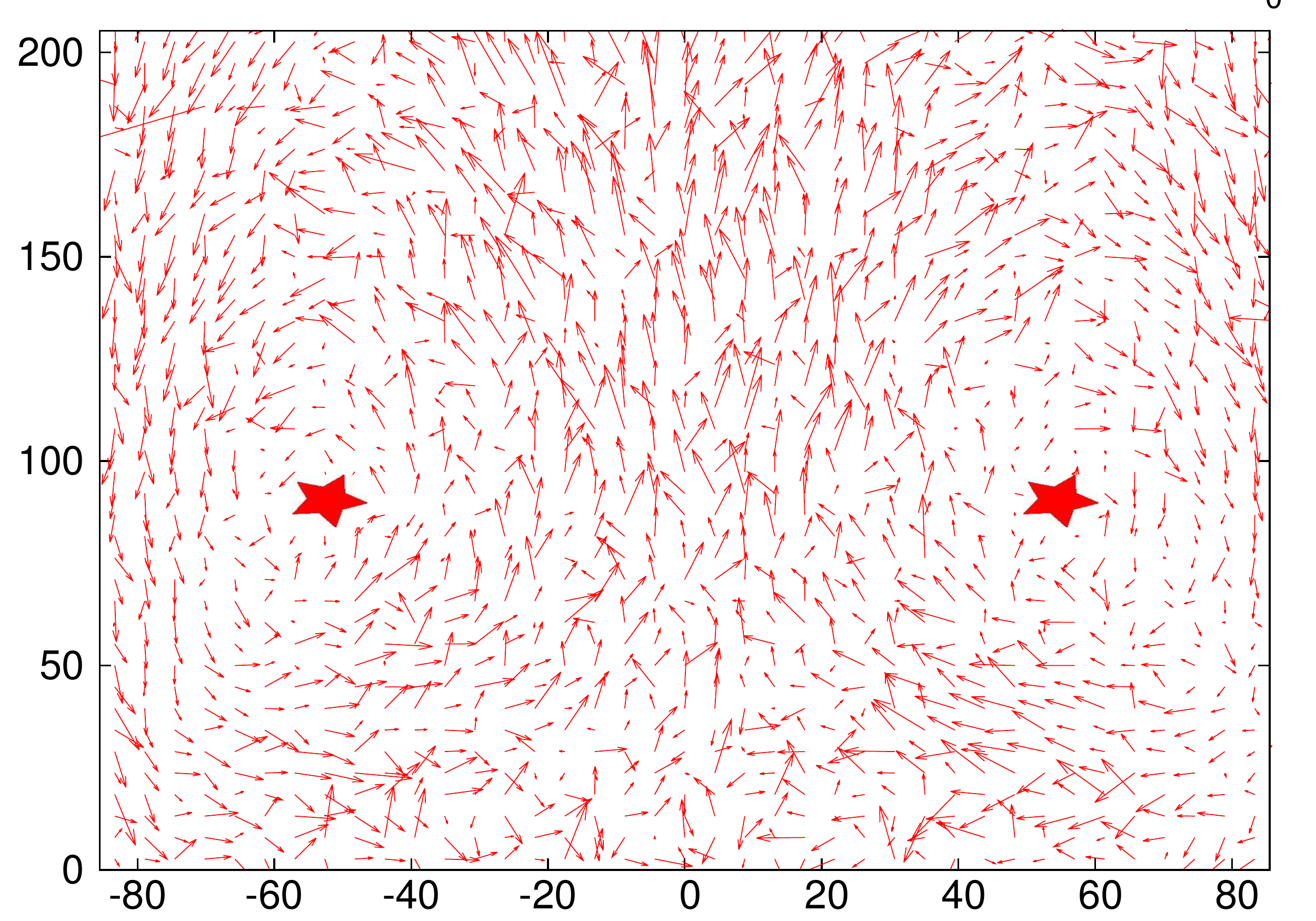}
  \includegraphics[width=3.5cm,height=4.0cm]{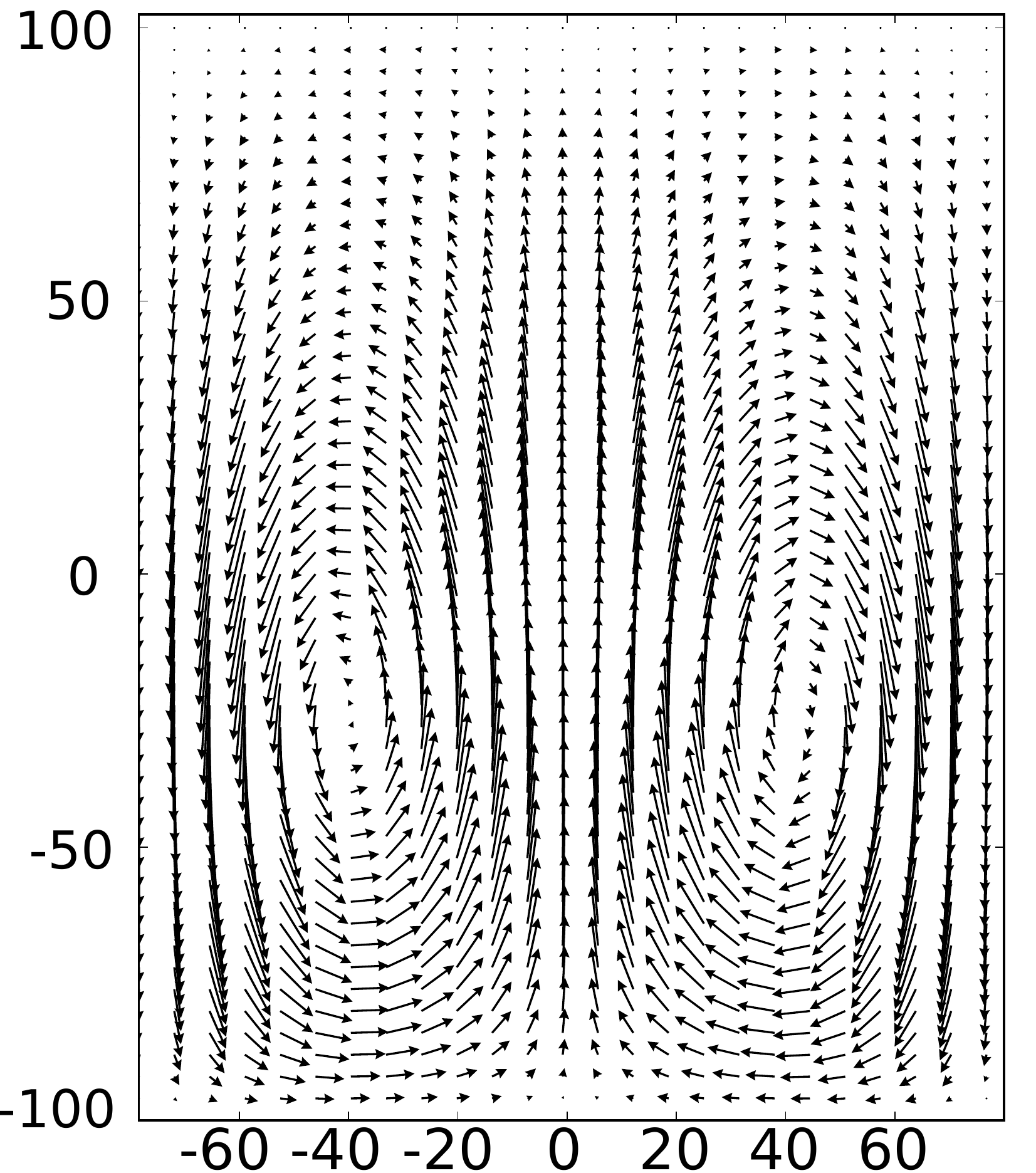}\\ \vspace{0.25cm}
 \includegraphics[width=12.75cm,height=3.5cm]{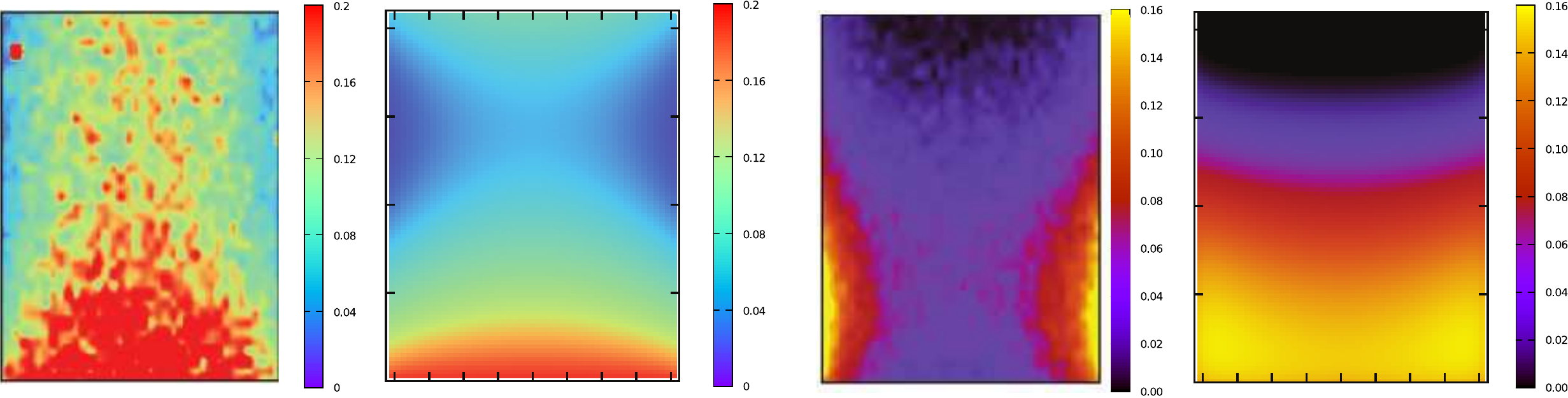}
 \caption{Fields from experiments and theory (left \& right panels respectively, for each
    pair of panels for the corresponding field), for a system without a top wall (open system). Length unit:
    particles diameter ($\sigma=1$ mm). $g=0.016\times 9.8~\mathrm{m/s^2}$. Top
    row: flow field ($\boldsymbol{u}=\boldsymbol{0}$). Bottom row, left half: the corresponding temperature $T/m$
    for experiment (first panel from the left) and theory (second panel from the
    left). Bottom row, right half: packing fraction fields $\nu=n \pi \sigma^2/4$ for
    experiment (left) and theory (right). Black stands for lower and red for higher field value. $0<T\le
   0.2mv_0^2$, $(v_0=370~\mathrm{mm/s}$); $0.02\times 10^{-2}<\nu\le0.15\times 10^{-2}$, mean packing
   fraction: $\nu = 0.049\times 10^{-2}$. Density color bars are in percentage
   units. Experiments: with $f=45$ Hz, $A=1.85$ mm, $N=300$ (total number of particles). Theory: coefficient of restitution $\alpha=0.9$ (both for particle-particle and wall-particle collisions).}
  \label{theory-exp}
  \end{center}
  \end{figure}

\begin{figure}
  \begin{center}
  \includegraphics[width=3.5cm,height=4.1cm]{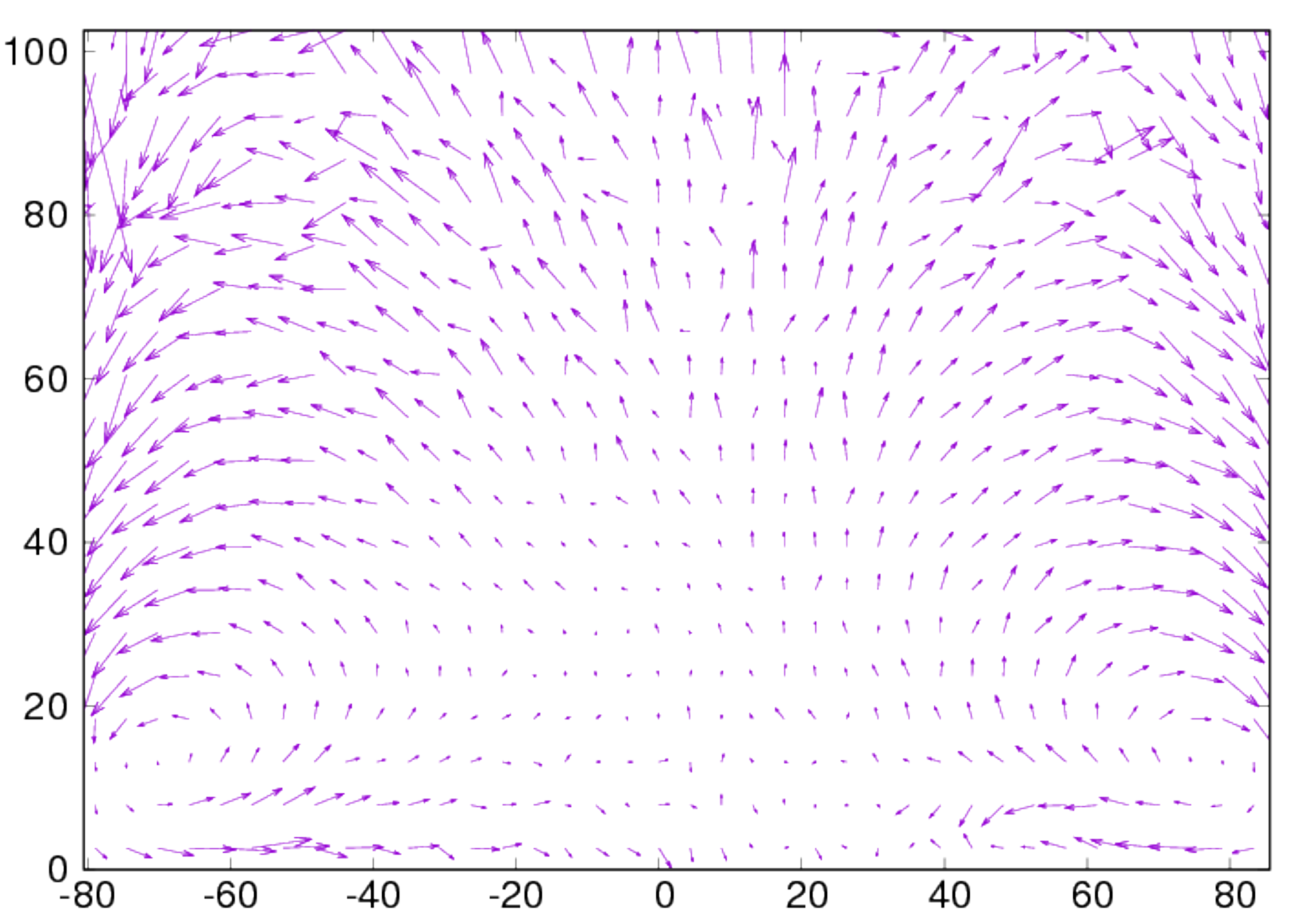}  \includegraphics[width=3.5cm,height=4.0cm]{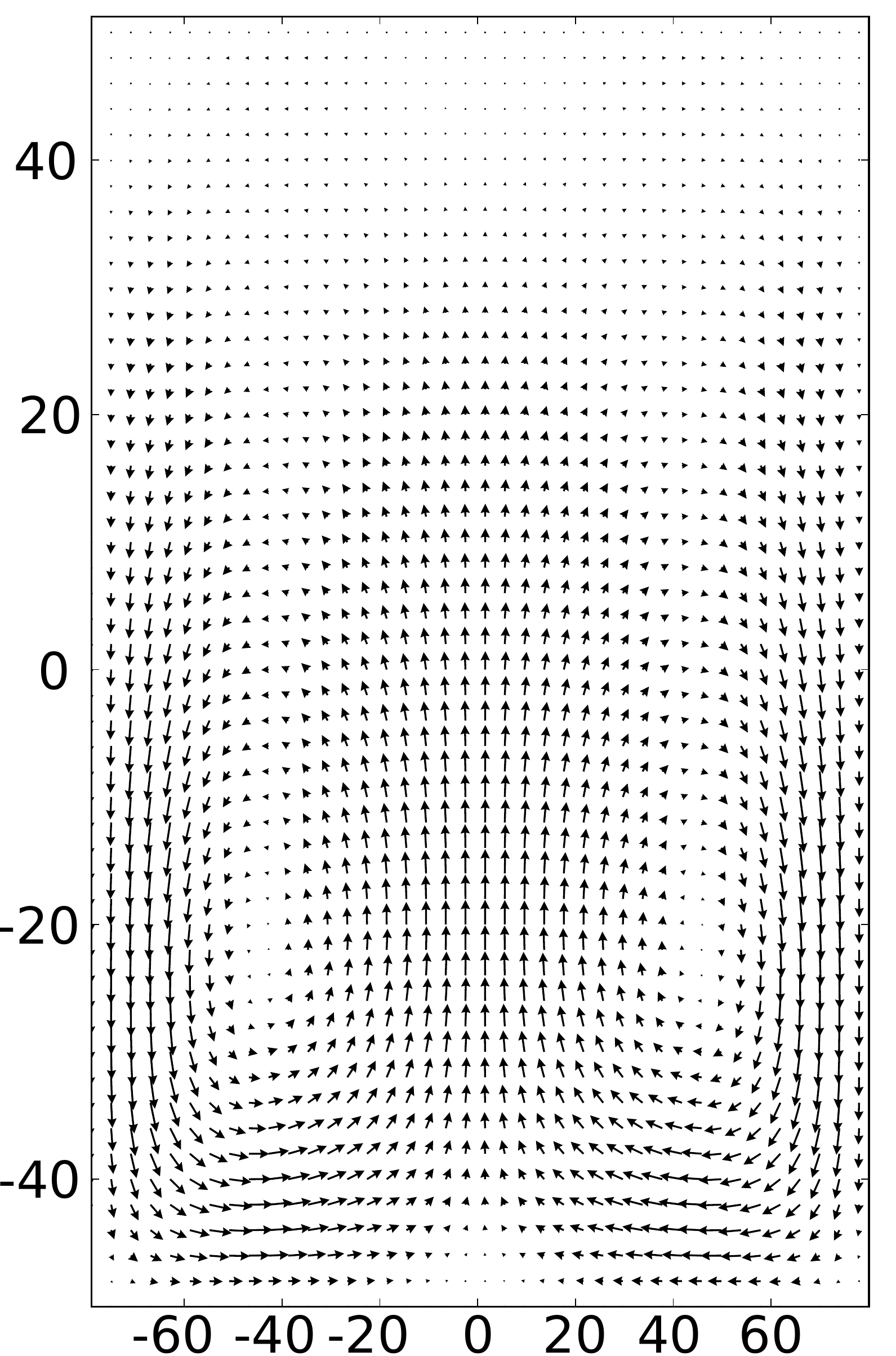}\\
  \includegraphics[width=2.75cm,height=3.25cm]{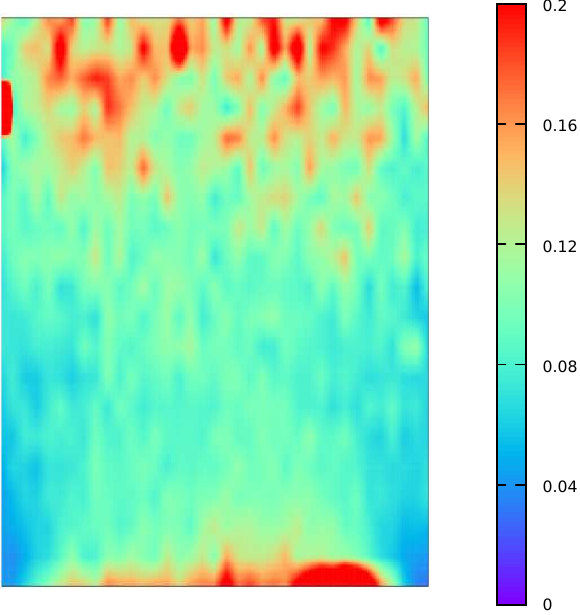}
  \includegraphics[width=2.75cm,height=3.25cm]{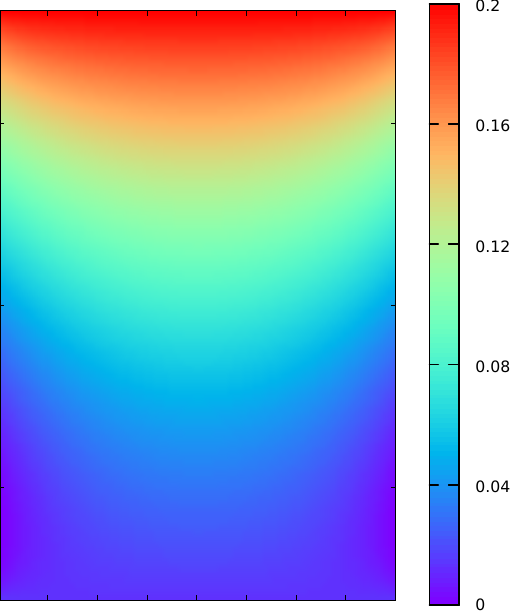}
  \quad
  \includegraphics[width=2.75cm,height=3.25cm]{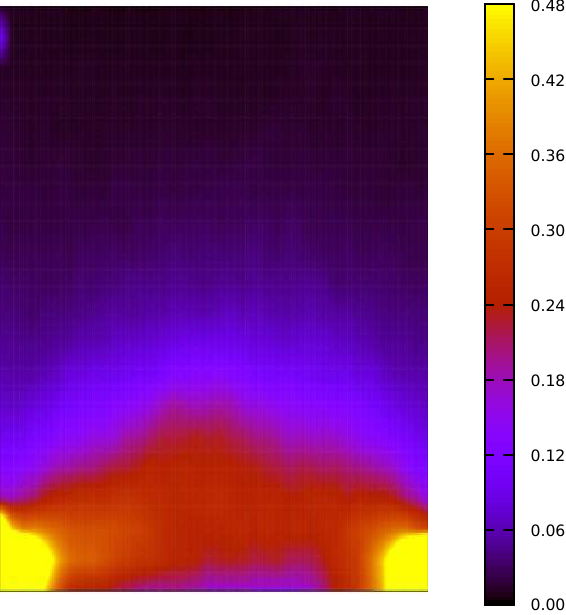}
 \includegraphics[width=2.75cm,height=3.25cm]{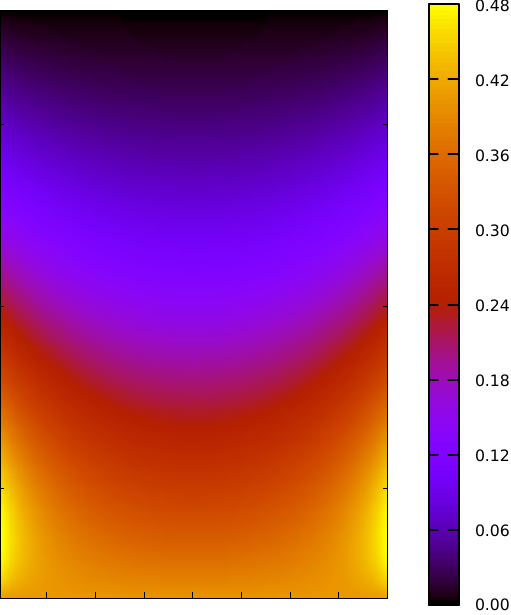}
  \caption{Fields from experiments and theory (left \& right panels, for each
    pair of panels for the corresponding field,
    respectively) for a system without a top wall (open system). Length unit:
    particles diameter ($\sigma=1$ mm). $g=0.016\times 9.8~\mathrm{m/s^2}$. Top
    row: flow field ($\boldsymbol{u}=\boldsymbol{0}$). Bottom row, left half: the corresponding temperature $T/m$ for
    experiment and theory (right). Bottom row, right half: packing fraction
    $\nu=n \pi \sigma^2/4$ fields for experiment (left) and theory (right). Black
    stands for lower and red for higher field value. $0<T\le 0.2mv_0^2$,
    $(v_0=370~\mathrm{mm/s}$); $0.02\times 10^{-2}<\nu\le0.46\times 10^{-2}$,
    mean packing fraction: $\nu=0.15\times 10 ^{-2}$. Density color bars are in
    percentage units. Experiments: with $f=45$ Hz, $A=1.85$ mm, $N=1000$ (total
    number of particles). Theory: coefficient of restitution $\alpha=0.9$ (both
    for particle-particle and wall-particle collisions).}
  \label{theory-exp-2}
  \end{center}
  \end{figure}

 In figures \ref{theory-exp} and \ref{theory-exp-2}, that correspond to open
 systems (i.e., without a top wall) we see a comparison between theory and
 experimental results, for $g=0.016~g_0$ (with $g_0=9.8~\mathrm{m/s^2}$) in the
 cases of $N=300$ and $N=1000$ particles in the experimental set up
 respectively. In all cases, we may consider the system as dilute, since the
 local packing fraction $\nu = n\pi\sigma^2/4$ is never greater than,
 approximately, $\nu=0.5\times 10^{-2}$. As we can see, the agreement is good
 for the flow field, and more qualitative for the temperature and density
 fields. In the case of Figure~\ref{theory-exp}, there is some disagreement
 between theory and experiment for both temperature and density, which be
 partially explained by the fact that the Knudsen number is not small
 ($\mathrm{Kn}\sim 10^{-1}$) and therefore, there might be non-Newtonian effects
 in the experiments that are not taken into account in our theory. However, it
 is clear  in both cases that the cold fluid regions next to the lateral walls
 are adjacent to the convective
 cell centers. We also checked that when dissipation at the lateral walls is
 switched off, no DLW convection appears out of the theoretical solution. In
 this way, we  may conclude that the convection mechanism appears as a
 consequence of the combined action of two perpendicular gradients: the density
 (and thermal) gradient due to the action of gravity and the horizontal gradient
 due to energy dissipation by the lateral walls.
 
Let us point out that our theory does not take into account all of the
ingredients and details that are present in previous experiments on DLWC in
granular systems \citep*{PGVP16,WHP01,WRP13}, such as plate-particle friction
and/or sliding, or dynamical effects derived from particle sphericity (just to
put two examples), etc. Moreover, as noticed in previous works, there is a
tendency do volume convection disappearance for not so small Knudsen numbers \citep{PGVP16,AA16}. For all this, a comparison with our previous
experimental results \cite{PGVP16} and the experimental results by others
\citep{WHP01,EMea10,WRP13} should be regarded as qualitative, not
quantitavive. The advantage of this procedure us that, because it is
reduced to the essentials, we have finally been able to identify the key
ingredients that produce the DLWC in the granular gas.

Furthermore, theoretical procedures (simulations, hydrodynamics) allow
us to separate the two sources of dissipation and make ``ideal''
assumptions. For instance, we can switch off internal dissipation (and reproduce
the molecular fluid limit case) while retaining dissipation
at the walls.


 \section{Conclusions}

We discuss in this work the theory framework for a previously observed
experimental phenomenon of granular convection
\citep*{WHP01,WHP01E,RSGC05,EMea10,WRP13,PGVP16}. This convection appears
automatically; i.e., occurs at arbitrary \Ray~ and, in close analogy to the
convection in molecular fluids
induced by cold sidewalls \citep{HW77}, we have concluded that the new granular
convection is also induced by dissipative vertical
sidewalls and a gravity field. We denote it as DLW convection. We have built a
granular hydrodynamic theoretical framework that explains the physical origin of this  
convection, in the context of the Boussinesq equations for the granular gas. To
our knowledge, it is the first time that a Boussinesq-type approach is used for
granular convection. We found that our
theory also explains the main features of the experimental observations.
That is, the DLW convection displays in all cases only one convection cell per
dissipative wall, the width of this wall being increased when convection intensity
increases. Moreover, the DLW convection intensity is enhanced by increasing gravity
acceleration, and/or wall dissipation. Conversely, it is decreased by increasing
bottom wall temperature (at fixed gravity and wall dissipation).  

Notice that sidewalls are inherently inelastic in granular fluid experimental
systems. Thus, the DLW convection is always present at experimental level and
our results imply that the classical volume thermal convection in granular gases
does not appear in experimental systems out of a hydrostatic state. Instead, it
develops as a secondary instability out of the DLWC state. Therefore, more
theory work would be needed in general to correctly describe the experimental
instability criteria for the volume thermal convection in granular gases. This
implies also that accuracy of previous hydrodynamic theory for the volume
thermal convection could be seemingly improved if the prior existence of the DLW
is taken into account. 

Finally, our present work constitutes another strong evidence that steady granular flows can be correctly described with a standard hydrodynamic theory \citep{P15} \citep[see also the recent work by][with results in the same line]{VL17}.

F. V. R. acknowledges support from grants ”Salvador
Madariaga” No. PRX14/00137, FIS2016-76359-P (both from
Spanish Government) and  No.GR15104 (Regional
Government of Extremadura, Spain). Use of computing facilities from the Extremadura Research Centre for Advanced Technologies (CETA-CIEMAT), partially financed by the ERDF is also acknowledged.

\bibliographystyle{jfm}
\bibliography{Bc_DLW}

\end{document}